\documentclass[
floatfix,
aps,
pra,
amsmath,
twocolumn,
superscriptaddress
]{revtex4-2}

\pdfoutput=1

\usepackage{graphicx}
\usepackage{wrapfig}
\usepackage{amsmath}
\usepackage{amssymb}
\usepackage{bbm}
\usepackage{hyperref}
\usepackage{soul}
\usepackage{xcolor}
\usepackage{bbold}
\usepackage{multirow}

\newcommand{\ket}[1]{|{#1}\rangle}
\newcommand{\bra}[1]{\langle{#1}|}

\begin{document}

\title{CNOT gates for fluxonium qubits via selective darkening of transitions}

\author{Konstantin N. Nesterov}
\email{konstantin@bleximo.com}
\affiliation{Department of Physics  and Wisconsin Quantum Institute, University of Wisconsin-Madison, Madison, WI 53706, USA}
\affiliation{Bleximo Corp., Berkeley, CA 94710, USA}

\author{Chen Wang}
\affiliation{Department of Physics, University of Massachusetts-Amherst, Amherst, MA, 01003, USA}

\author{Vladimir E. Manucharyan}
\affiliation{Department of Physics,
University of Maryland, College Park, MD 20742, USA}

\author{Maxim G. Vavilov}
\affiliation{Department of Physics and Wisconsin Quantum Institute, University of Wisconsin-Madison, Madison, WI 53706, USA}

\date{\today}

\begin{abstract}
We analyze the cross-resonance effect for fluxonium circuits and investigate a two-qubit gate scheme based on selective darkening of a  transition.  In this approach, two microwave pulses at the frequency of the target qubit are applied simultaneously with a proper ratio between their amplitudes to achieve a 
controlled-\textsc{not} operation. We 
study in detail coherent gate dynamics and calculate gate error. 
With nonunitary effects accounted for, we demonstrate that gate error below $10^{-4}$ is possible for realistic hardware parameters. This number is facilitated by long coherence times of computational transitions and strong anharmonicity of fluxoniums, which easily prevents excitation to 
higher excited states during the gate microwave drive. 
\end{abstract}

\maketitle

\section{Introduction}\label{sec-introduction}

Two-qubit gates with very low error rates are essential for making a practical quantum processor. In superconducting architectures~\cite{Devoret2013, Wendin2017, Kjaergaard2020_review},  such operations with fidelity approaching 99.9\% have already been demonstrated~\cite{Sung2021, Kandala2021,  Negirneac2021, Wei2022}. In the family of two-qubit gates utilizing microwave control, one of the most successful and popular gate schemes is based on the cross-resonance (CR) effect, which results in the controlled-\textsc{not} (\textsc{cnot}) operation up to local single-qubit rotations~\cite{Paraoanu2006, Rigetti2010}. This gate is activated by driving the control qubit at the frequency of the target qubit (CR drive).
It has been implemented experimentally multiple times, including experiments with ordinary flux qubits~\cite{deGroot2010}, capacitively shunted flux qubits~\cite{Chow2011}, and  transmons~\cite{Chow2012, Corcoles2013, Takita2016, Sheldon2016b, Jurcevic2021, Kandala2021, Wei2022}. 
 These experiments are accompanied by several theoretical publications that discuss the gate and techniques of its improvement in great detail~\cite{deGroot2012, Kirchhoff2018, Tripathi2019, Magesan2020, Sundaresan2020, Malekakhlagh2020, Malekakhlagh2022}.

Despite being  a common superconducting qubit, the transmon has a  drawback of having a weakly anharmonic energy spectrum, which generally complicates qubit control. A promising alternative to the transmon is the fluxonium circuit with its strongly anharmonic spectrum and long coherence time of the low-frequency transition between the ground and first excited states~\cite{Manucharyan2007, Nguyen2019, Somoroff2021}.  One of the microwave-activated two-qubit gate schemes that has been implemented experimentally with fluxonium qubits is based on driving in proximity with transitions leading to higher (noncomputational) excited states of the two-qubit spectrum~\cite{Ficheux2021, Xiong2022}. Such gate operations are facilitated by several-gigahertz transmon-like frequencies of those transitions, which results in stronger interactions between noncomputational levels in comparison to interactions between computational levels~\cite{Nesterov2018}. A gate scheme based on populating higher excited states was also suggested for heavy fluxoniums~\cite{Abdelhafez2020}. To fully benefit from a long coherence time of the main qubit transitions of fluxoniums, the population, however, should remain in the computational subspace during the gate operation. This reasoning has been used in a recent proposal of a two-qubit gate based on activating a two-photon transition by a strong microwave drive~\cite{Nesterov2021}. Because the fluxonium spectrum is anharmonic, a strong drive amplitude in that proposal does not cause significant leakage to noncomputational levels, and gate fidelity is therefore not spoiled by shorter coherence times of higher excited states. Other methods to remain in the computational subspace are to implement flux-tunable gates~\cite{Zhang2021, Chen2022a, Bao2022} or tunable couplers~\cite{Moskalenko2021}, although these schemes may incur first-order flux noise and require added hardware complexity. 

In this paper, we explore theoretically the CR gate for fluxoniums and therefore continue studying gate schemes that benefit in full from long coherence time of computational transitions. 
We focus on the selective-darkening (SD) variant of the CR gate, which was  studied both experimentally~\cite{deGroot2010} and theoretically~\cite{deGroot2012}  for flux qubits  and has been lately used for transmons under the name of  direct \textsc{cnot}~\cite{Jurcevic2021, Kandala2021, Malekakhlagh2022, Wei2022}. In addition,  more recently, the SD variant was demonstrated for fluxoinum circuits~\cite{Dogan2022}. In this scheme, a strong CR drive of the control qubit is accompanied by a weak drive of the target qubit that is chosen to cancel rotation of the target qubit when the control qubit is in its ground state. When the control qubit is in its excited state, the target qubit rotates, which, with proper calibration, results in \textsc{cnot} operation. Using the language of effective Hamiltonian models~\cite{Paraoanu2006, Magesan2020}, these two tones result in effective $ZX$ and $IX$ terms of equal magnitudes but opposite signs. The primary effect of the CR drive without the second tone is to produce some unbalanced combination of $ZX$ and $IX$ terms, which requires an additional single-qubit rotation of the target qubit to achieve \textsc{cnot}. This basic scheme can be improved by using an echo sequence of two CR drives of opposite signs and of additional $\pi$ rotations of the control qubit, which eliminates various spurious terms in the effective Hamiltonian such as $ZI$ and $ZZ$ and makes the operation insensitive to the low-frequency amplitude noise~\cite{Corcoles2013, Takita2016, Sheldon2016b, Malekakhlagh2020, Sundaresan2020}. 

Nevertheless, the SD approach, which produces a direct \textsc{cnot} operation without using an echo sequence and additional single-qubit rotations, results in faster gates and has been demonstrated experimentally to work well for transmons~\cite{Jurcevic2021, Kandala2021, Wei2022}. We illustrate via simulations that such a scheme produces high-fidelity and fast gates for fluxoniums as well, which holds even when $ZZ$ crosstalk is relatively strong. We show that this operation is facilitated by the structure of interaction matrix elements, which are enhanced for transitions to higher energy levels, eventually allowing one to achieve reasonably fast speeds for the proposed two-qubit gate.  The operation requires single-qubit $Z$ rotations, which can be done instantly in  software~\cite{McKay2017}. 
 For realistic fluxonium parameters with qubit lifetimes of 500 $\mu$s and 50 $\mu$s of the first and second excited states, we find greater than $99.99\%$ coherent fidelity for gate duration of 50 ns without using any advanced pulse shaping. In experiment of Ref.~\cite{Dogan2022}, the gate error was shown to be about 0.5\% with qubits having short coherence times in the range of 10-20 $\mu$s.  The incoherent error is dominated by lifetime of the computational subspace: even when the relaxation time of the second excited states in simulations is only 1 $\mu$s, its contribution to this error is below 0.1\%.

The outline of the paper is as follows. In Sec.~\ref{sec-model}, we review the model of two coupled fluxoniums, the gate concept, discuss in detail the structure of charge matrix elements relevant for the gate operation, and elaborate on fundamental limitations on gate rate. In Sec.~\ref{sec-fidelity}, we discuss our simulations of the unitary dynamics, coherent error budget, and the reduction of gate fidelity by nonunitary processes. Finally, we conclude in Sec.~\ref{sec-conclusions}.

\begin{figure}[t]
\includegraphics[width=\columnwidth]{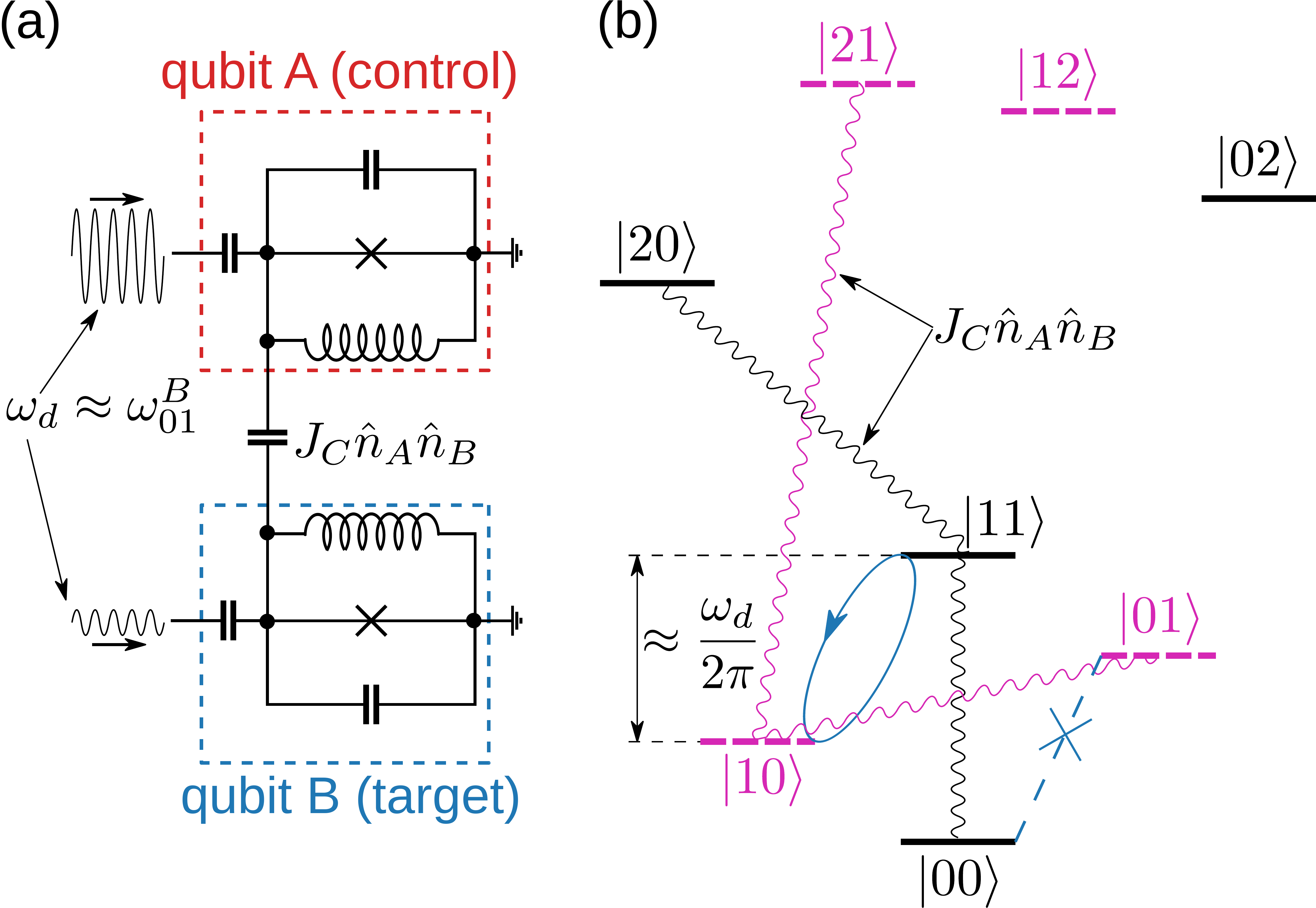}\caption{
Circuit diagram (a) and energy levels (b) of two capacitively coupled fluxonium qubits $A$ and $B$ driven by two local microwave fields with frequency $\omega_d$.  When $\omega_d \approx \omega_{01}^B$, a strong drive of qubit $A$ and a properly chosen much weaker drive of qubit $B$ realize a controlled-$U$ gate operation on qubit $B$ (target), so qubit $B$ is rotated only when qubit $A$ (control) is in $\ket{1}$.  Wavy lines in panel (b) illustrate  pairs of hybridized two-qubit levels that give dominant contributions to $\langle 10 |\hat{n}_A| 11\rangle$.
}\label{Fig-introduction}
\end{figure}

\section{Model and gate concept}\label{sec-model}

\subsection{Interacting fluxoniums}

We consider the system of two capacitively coupled fluxonium circuits in the presence of two local microwave drives, which is shown schematically in Fig.~\ref{Fig-introduction}(a). Without the loss of generality, we assume that qubits $A$ and $B$ are the control and target qubits, respectively, so the local drive applied to qubit $A$ is the CR drive, whose amplitude is larger than the amplitude of the second drive. Both drives are applied at the same frequency $\omega_d\approx \omega_{01}^B$,  where $\omega^\alpha_{kl}$ is the frequency of the single-qubit transition $\ket{k_\alpha}-\ket{l_\alpha}$ between two eigenstates of qubit $\alpha$ ($\alpha = A, B$). 

We model this system by the Hamiltonian
\begin{equation}\label{Hamiltonian-two-qubit}
\hat{H} = \hat{H}^{(0)}_{A} + \hat{H}^{(0)}_B + \hat{V} + \hat{H}_{\rm drive}\,,
\end{equation}
where the first two terms describe individual qubits and are given by 
\begin{equation}\label{Hamiltonian-fluxonium}
 \hat{H}_{\alpha}^{(0)} = 4E_{C,\alpha} \hat{n}_\alpha^2 + \frac 12 E_{L,\alpha} \hat{\varphi}_\alpha^2 - E_{J,\alpha} \cos(\hat{\varphi}_\alpha - \phi_{\rm ext,\alpha})\,
\end{equation}
with $\alpha = A, B$. Here $\hat{\varphi}_\alpha$ and $\hat{n}_\alpha$ are the dimensionless position-like flux operator and momentum-like charge operator with the  commutation relation $[\hat{\varphi}_\alpha, \hat{n}_{\alpha'}] = i\delta_{\alpha\alpha'}$.
The third term in Eq.~(\ref{Hamiltonian-fluxonium}) describes capacitive interaction according to
\begin{equation}\label{interaction-charge}
 \hat{V} = J_C \hat{n}_A \hat{n}_B\,.
\end{equation}
Parameters of Eqs.~\eqref{Hamiltonian-fluxonium} and \eqref{interaction-charge} are discussed in detail in Refs.~\cite{Nesterov2018, Nesterov2021}.
Here we briefly note that each qubit is characterized by its charging ($E_{C, \alpha}$), inductive ($E_{L, \alpha}$), and Josephson ($E_{J,\alpha}$) energies as well as the  dimensionless variable $\phi_{{\rm ext}, \alpha}$, which is proportional to the external magnetic flux threading the loop formed by the superinductor and Josephson junction.  The value of $\phi_{{\rm ext}, \alpha}$ is tunable \emph{in situ}, and, similarly to other microwave-activated schemes, here we consider fluxoniums permanently parked at their sweet spots of maximal coherence at $\phi_{{\rm ext}, \alpha} = \pi$~\cite{Nguyen2019, Somoroff2021}. The interaction strength $J_C$ in Eq.~(\ref{interaction-charge}) is determined by the mutual capacitance and individual capacitances of the two qubits~\cite{Nesterov2018, Nesterov2021}. 
Finally, the last term in Eq.~\eqref{Hamiltonian-two-qubit} describes the coupling to two external microwave drives of frequency $\omega_d$:
\begin{equation}\label{drive}
 \hat{H}_{\rm drive} =2\hbar f(t)\cos(\omega_d t) \left(\hat{n}_A +  \eta\hat{n}_B\right) \,.
\end{equation}
Here $\hbar = h/2\pi$ is the reduced Planck constant, $f(t)$ is the time-dependent field envelope, and $\eta$ captures the combined effect of different drive amplitudes and of classical crosstalk. We emphasize that because of this crosstalk in a realistic system, each local drive couples to both qubits, so $\eta$ is not simply the ratio of two drive amplitudes. However, if the ratio of two amplitudes can be tuned, the value of $\eta$ can be tuned as well.

\subsection{Gate concept}

An essential condition of any CR scheme is the dependence of the drive matrix element for target-qubit transitions on the state of the control qubit. Let $\ket{kl}$ be the dressed two-qubit eigenstate of Hamiltonian~\eqref{Hamiltonian-two-qubit} at $\hat{H}_{\rm drive}=0$ corresponding to the noninteracting tensor-product state $\ket{k_A}\ket{l_B}$. Then, for our choice of the control and target qubits, each target-qubit transition is between states $\ket{k0}$ and $\ket{k1}$ for some $k$. Therefore, the essential CR condition implies
\begin{equation}\label{drive_inequality}
    \langle 00|\hat{H}_{\rm drive}| 01\rangle \ne \langle 10|\hat{H}_{\rm drive}| 11\rangle\,.
\end{equation} 
This way, the Bloch-sphere trajectory of qubit $B$ in the presence of $\hat{H}_{\rm drive}\ne 0$ depends on the state of qubit $A$. 
 When $J_C=0$, we find that the inequality \eqref{drive_inequality} is violated since both sides reduce to the same value determined by the single-qubit charge matrix element $\langle 0_B|\hat{n}_B|1_B\rangle$ for the target qubit. When $J_C\ne 0$, two-qubit bare states $\ket{k_A}\ket{l_B}$ hybridize to form dressed states $\ket{kl}$. Because of this hybridization, $\bra{k0} \hat{H}_{\rm drive} \ket{k1}$ acquires a $k$-dependent correction coming from corrections to both $\bra{k0} \hat{n}_{A} \ket{k1}$ and $\bra{k0} \hat{n}_{B} \ket{k1}$. 

The SD condition requires one of the transition matrix elements of $\hat{H}_{\rm drive}$ to vanish.  To be specific, we take
\begin{equation}\label{sd_condition}
    \langle 00|\hat{H}_{\rm drive}| 01\rangle = 0\,,
\end{equation}
which, together with inequality \eqref{drive_inequality}, implies that only $\ket{10}-\ket{11}$ transition is activated, while $\ket{00}-\ket{01}$ transition is made forbidden, see Fig.~\ref{Fig-introduction}(b).
Using Eq.~\eqref{drive}, we find that the SD condition~\eqref{sd_condition} is equivalent to
\begin{equation}\label{eta}
    \eta = - \frac{\bra{00}\hat{n}_A \ket{01} }{  \bra{00}\hat{n}_B \ket{01}}\,.
\end{equation}
The resonance Rabi frequency for the $\ket{10}-\ket{11}$ transition for the continuous drive $f(t)={\rm const.}$ in Eq.~\eqref{drive} is then given by
\begin{equation}\label{Omega_10_11_exact}
    \Omega_{10-11} = 2f\left| \bra{10}\hat{n}_A \ket{11} - \bra{00}\hat{n}_A \ket{01} \frac{\bra{10}\hat{n}_B \ket{11}}{\bra{00}\hat{n}_B \ket{01}}\right|\,.
\end{equation}
The \textsc{cnot} gate duration is given by half period of Rabi oscillations: $t_{\rm gate} = \pi / \Omega_{10-11}$.

We further refer to the two-qubit charge matrix elements of the type $\bra{k0} \hat{n}_{A} \ket{k1}$ as cross matrix elements; they are zero at $J_C=0$. Matrix elements of the second type, $\bra{k0} \hat{n}_{B} \ket{k1}$, are referred to as direct matrix elements; they reduce to single-qubit values at $J_C=0$.
The first nonvanishing correction to the cross matrix elements  is linear in $J_C$, while it is only quadratic for the direct matrix elements  because of the  parity selection rules for the charge operators at half flux quantum~\cite{Nesterov2018, Nesterov2021}. Therefore, to linear order in $J_C$, we find
\begin{equation}\label{Omega_10_11_nAonly}
    \Omega_{10-11} = 2f \left|\bra{10}\hat{n}_A \ket{11} - \bra{00}\hat{n}_A \ket{01} \right| + O\left(J_C^2\right)\,.
\end{equation}
By analogy, if qubit $A$ is chosen as the target qubit and qubit $B$ is the control, we find
\begin{equation}\label{Omega_01_11_nBonly}
    \Omega_{01-11} = 2f \left|\bra{01}\hat{n}_B \ket{11} - \bra{00}\hat{n}_B \ket{10} \right| + O\left(J_C^2\right)\,.
\end{equation}
Therefore, to linear order in $J_C$, Rabi rates for the CR effect are determined by the cross matrix elements.
We calculate the values of both cross and direct charge matrix elements in the next section.

\subsection{Charge matrix elements}\label{sec-charge-mel}

\begin{table*}[t]
    \centering
    \begin{tabular}{cccccccccc}
    \hline\hline
    \multirow{2}{*}{Qubit} 
         & $E_{L,\alpha}/h$ & $E_{C,\alpha}/h$ & $E_{J,\alpha}/h$ & $\omega^\alpha_{01}/2\pi$ & $\omega^\alpha_{12}/2\pi$
         & $\omega^\alpha_{03}/2\pi$ 
         & 
         \multirow{2}{*}{$|\langle 0_\alpha |\hat{n}_\alpha |1_\alpha\rangle |$} & 
         \multirow{2}{*}{$|\langle 1_\alpha |\hat{n}_\alpha |2_\alpha\rangle |$}
         & 
         \multirow{2}{*}{$|\langle 0_\alpha |\hat{n}_\alpha |3_\alpha\rangle |$}
         \\
          & (GHz) & (GHz) & (GHz) & (GHz) & (GHz) & (GHz) &  & &\\
         \hline
         $A$ & 1.09  & 1.06  & 4.62  & 0.53  & 3.80  
         & 7.03  & 0.14 & 0.58 & 0.41 
         \\
         $B$ & 1.88  & 1.03  & 5.05  & 1.02  & 3.75  
         & 8.25  & 0.22 & 0.63 & 0.32
         \\
         \hline\hline
    \end{tabular}
    \caption{Fluxonium parameters used for numerical simulations.}
    \label{Table-params}
\end{table*}

\begin{figure}[t]
\includegraphics[width=\columnwidth]{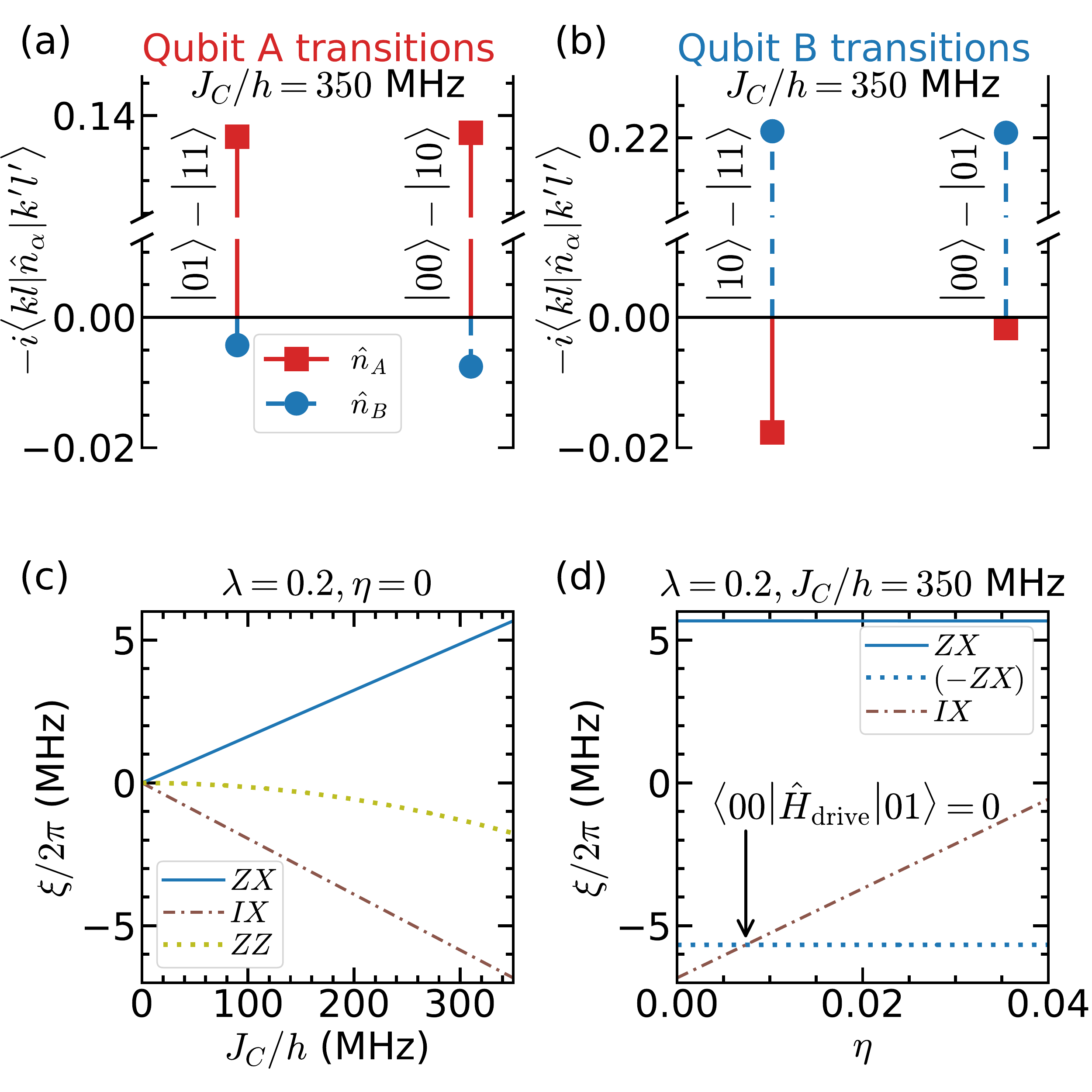}\caption{(a), (b) Matrix elements of qubit charge operators $\hat{n}_A$ (red squares) and $\hat{n}_B$ (blue circles) for various computational transitions calculated for parameters of Table~\ref{Table-params} at $J_C/h = 350$ MHz. The cross matrix elements, e.g., $\bra{10}\hat{n}_A \ket{11}$,  give the dominant contribution to the CR effect. (c) The rates of effective $ZX$ (blue solid line), $IX$ (brown dash-dot line), and $ZZ$ (cyan dotted line) interactions calculated using Eqs.~\eqref{xi_zx}-\eqref{xi_zz} vs $J_C$ for $\lambda=0.2$ [see Eq.~\eqref{lambda}] and $\eta=0$. (d) The rates of $ZX$, negative $ZX$ (blue dotted line), and $IX$ interactions vs $\eta$ at $\lambda=0.2$ and $J_C/h=350$ MHz.  }\label{Fig-matrels}
\end{figure}

For quantitative analysis, we use realistic hardware parameters shown in Table~\ref{Table-params} with variable interaction strength $J_C$. Single-qubit frequencies for these parameters are in the 500-1000 MHz range, which ensures that  $n_{01}^\alpha$, where $n_{kl}^\alpha = -i\bra{k_\alpha}\hat{n}_\alpha\ket{l_\alpha}$, is not very small to allow a sufficient  hybridization of computational levels with relevant states both inside and outside of the computational subspace. In comparison, devices with single-qubit transition frequencies in the 100-200 MHz range result in $n_{01}^\alpha \lesssim 0.05$, making gate operations based on hybridization of computational levels more challenging to implement. Such devices suit better for gate schemes involving higher noncomputational levels because of larger $n_{12}^\alpha$ and $n_{03}^\alpha$.
In particular, single-qubit transition frequencies in the controlled-$Z$ gate realization of Ref.~\cite{Ficheux2021}, which was based on driving in proximity with the $\ket{11}-\ket{21}$ transition, were only 70 and 130 MHz.  

In Figs.~\ref{Fig-matrels}(a) and \ref{Fig-matrels}(b), we show both direct and cross  matrix elements of $\hat{n}_A$ and $\hat{n}_B$ calculated numerically 
for parameters of Table~\ref{Table-params} and $J_C/h = 350$ MHz. We make three main observations. 
First, we notice that direct charge matrix elements  are large in comparison to cross matrix elements, e.g., $|\bra{00} \hat{n}_A \ket{10}|\gg |\bra{00} \hat{n}_B \ket{10}|$ in Fig.~\ref{Fig-matrels}(a), since the cross matrix elements become nonzero only due to interaction-induced corrections, i.e., $\bra{00} \hat{n}_B \ket{10} = 0$ when $J_C=0$.
Second, we notice, however, that the \emph{change} of the matrix element with the qubit state changing is greater for cross matrix elements, e.g., $|\bra{10} \hat{n}_B \ket{11} - \bra{00} \hat{n}_B\ket{01}| < |\bra{10} \hat{n}_A \ket{11} - \bra{00} \hat{n}_A\ket{01}|$ in Fig.~\ref{Fig-matrels}(b). This fact is in line with our previous reasoning that the interaction effects  are linear in $J_C$ in cross matrix elements and are quadratic in $J_C$ in direct matrix elements. 
Finally, we also observe that $\Omega_{10-11} > \Omega_{01-11}$ since $|\bra{10} \hat{n}_A \ket{11} - \bra{00} \hat{n}_A\ket{01}| > |\bra{01} \hat{n}_B \ket{11} - \bra{00} \hat{n}_B\ket{10}|$.  Therefore, we have chosen  qubit $B$ as the target qubit for parameters of Table~\ref{Table-params}.

We find that magnitudes of cross matrix elements are well explained by an approximation based on first-order perturbation theory that accounts only for contributions coming from computational levels and from $\ket{2k}$, $\ket{k2}$, $\ket{3k}$, and $\ket{k3}$, where $k=0, 1$. Analytic expressions for this approximation are derived in Appendix~\ref{sec-perturbation}. As an example, in Fig.~\ref{Fig-introduction}(b),  wavy lines show contributions to $\bra{10}\hat{n}_A\ket{11}$ coming from various pairs of hybridized levels. We emphasize that in comparison to true two-level systems, here couplings to higher levels such as that between $\ket{10}$ and $\ket{21}$ are relevant and cannot be ignored. 
In fact, the dominant contributions to $\bra{10}\hat{n}_A\ket{11}$ are coming from hybridization of $\ket{10}$ with $\ket{21}$ and of $\ket{11}$ with $\ket{20}$ rather than from hybridization of levels within the computational subspace. 
We also notice that because $\omega_{01}^A < \omega_{01}^B$, see Table~\ref{Table-params}, all four contributions to $\bra{10}\hat{n}_A\ket{11}$ interfere constructively, while, e.g., there is a destructive interference between contributions to $\bra{01}\hat{n}_B\ket{11}$. In addition, since charge matrix elements for the qubit transition approximately scale with frequency, $n_{01}^B$ is almost twice as large as $n_{01}^A$, which further increases contributions from hybridization of $\ket{10}$ and $\ket{21}$ and of $\ket{11}$ and $\ket{20}$ and eventually makes $\bra{10}\hat{n}_A\ket{11}$ the largest cross matrix element among four of them, see Figs.~\ref{Fig-matrels}(a) and \ref{Fig-matrels}(b).

\subsection{Effective Hamiltonian}

In this section, we use the language of effective Hamiltonian models~\cite{Paraoanu2006, Magesan2020} to give a complementary perspective on gate operation. The effective Hamiltonian is restricted to the computational subspace, has a block-diagonal structure with respect to control-qubit states, is written in the appropriate rotating frame of two qubits, and describes an effective interaction induced by a microwave drive in addition to a static $ZZ$ coupling. To focus on gate concept, we write it only to linear order in $f$, which yields
\begin{equation}\label{H_eff}
    \hat{H}_{\rm eff} = \frac{\xi_{ZX}}{2} \hat{\sigma}_z^A \otimes \hat{\sigma}_x^B + \frac{\xi_{IX}}{2} \hat{\sigma}_0^A \otimes \hat{\sigma}_x^B +   \frac{\xi_{ZZ}}{4} \hat{\sigma}_z^A \otimes \hat{\sigma}_z^B\,.
\end{equation}
Here $\xi_{ZX}$, $\xi_{IX}$, and $\xi_{ZZ}$ are the rates of effective $ZX$, $IX$, and $ZZ$ interactions, $\hat{\sigma}_0^A$ is the identity $2\times 2$ matrix, and $\hat{\sigma}_{i}^\alpha$ with $i=x, z$ is the Pauli matrix for qubit $\alpha$. It is $ZX$ interaction that is essential for any gate based on the CR effect.  For the SD gates, particularly, $\xi_{ZX} = \pm \xi_{IX}$. 
We note that even though we use tensor-product notation in Eq.~\eqref{H_eff}, this Hamiltonian is written in the interacting (dressed) eigenbasis rather than in the basis of tensor-product noninteracting states.

To linear order in $f$, we find
\begin{subequations}
\begin{equation}\label{xi_zx}
    \begin{aligned}
       \xi_{ZX} = f &\left[\bra{00}\hat{n}_A\ket{01} - \bra{10}\hat{n}_A \ket{11} \right. \\
       & \left. + \eta \left(\bra{00}\hat{n}_B\ket{01} - \bra{10}\hat{n}_B \ket{11}\right) \right]\,,
    \end{aligned}
\end{equation}
\begin{equation}\label{xi_ix}
    \begin{aligned}
       \xi_{IX} = f &\left[\bra{00}\hat{n}_A\ket{01} + \bra{10}\hat{n}_A \ket{11} \right. \\
       & \left. + \eta \left(\bra{00}\hat{n}_B\ket{01} + \bra{10}\hat{n}_B \ket{11}\right) \right]\,,
    \end{aligned}
\end{equation}
and
\begin{equation}\label{xi_zz}
    \xi_{ZZ} = \omega_{10-11} - \omega_{00-01}\,.
\end{equation}
\end{subequations}
If higher-order terms are properly taken into account, $IX$ and $ZX$ rates saturate with increasing the drive amplitude $f$, $ZZ$ rates acquire small corrections, and two new terms - $ZI$ and $IZ$ - appear in Eq.~\eqref{H_eff}~\cite{Tripathi2019, Magesan2020}. The origin of the $ZI$ term is AC Stark effect due to an off-resonance drive of the control qubit. While this effect is formally quadratic in $f$, its magnitude is relatively large because it is of zeroth order in $J_C$, so the induced $ZI$ rate quickly becomes dominant in the effective Hamiltonian~\cite{Magesan2020}. In comparison, $IZ$ rate and $f$-dependent correction to $ZZ$ rate are of higher order in both $J_C$ and $f$ and are rather small~\cite{Magesan2020}.
A possibly large magnitude of $ZI$ rate is not relevant for gate operation as its effect can be easily absorbed into virtual single-qubit $Z$ rotations~\cite{McKay2017}.

In Fig.~\ref{Fig-matrels}(c), we plot $ZX$, $IX$, and $ZZ$ rates calculated using perturbative Eqs.~\eqref{xi_zx}-\eqref{xi_zz} as a function of the interaction strength $J_C$ assuming only qubit $A$ is driven, i.e., at $\eta=0$. The drive amplitude $f$ is chosen to correspond to $\lambda=0.2$, where $\lambda$ is the dimensionless drive amplitude for qubit $A$ defined according to
\begin{equation}\label{lambda}
    \lambda = \frac{\Omega_{A, 0}}{\Delta_{AB}}\,.
\end{equation}
Here $\Omega_{A, 0} = 2f n^A_{01}$ is the single-qubit resonance Rabi frequency for qubit $A$ and $\Delta_{AB} = \omega^B_{01} - \omega^A_{01}$ is the detuning between qubit frequencies. The amplitude $\lambda$ is a measure of the strength of the off-resonance Rabi oscillations of qubit $A$ during the CR pulse; their contrast is given by $\lambda^2 / (\lambda^2 + 1)$. Linear results of Eqs.~\eqref{xi_zx}-\eqref{xi_zz} are valid under the condition  $\lambda \ll 1$. 

Figure~\ref{Fig-matrels}(c) illustrates that the strengths of $IX$ and $ZX$ terms are comparable, which signifies the contribution of higher noncomputational levels into these rates. In comparison, in purely two-level models, $\xi_{IX}=0$~\cite{Chow2011, Magesan2020}. 

The SD variant of the CR scheme, Eq.~\ref{sd_condition}, implies that $\xi_{IX} + \xi_{ZX} = 0$, which can be achieved by varying the drive amplitude of the second pulse applied to qubit $B$ or parameter $\eta$. We illustrate this statement in Fig.~\ref{Fig-matrels}(d) by plotting $IX$, $ZX$, and negative $ZX$ rates vs $\eta$ for the same $\lambda$ as in Fig.~\ref{Fig-matrels}(c) and the same $J_C$ as in the top panels of Fig.~\ref{Fig-matrels}. We observe that $\xi_{ZX}$ is almost unaffected by changing $\eta$ because the difference between two direct matrix elements in Eq.~\ref{xi_zx} is negligible as discussed in Sec.~\ref{sec-charge-mel}. In comparison, $\xi_{IX}$ contains the sum of two direct matrix elements, so it strongly depends on $\eta$. In other words, a direct resonant drive of qubit $B$ induces its rotation irrespective of state of qubit $A$, which is characterized by a change in $\xi_{IX}$. 

\subsection{Speed limit}\label{sec_speed_limit}

In addition to activating controlled rotations of the target qubit, the CR drive \eqref{drive} is applied  off-resonantly to transitions of the control qubit, which contributes to coherent control errors and eventually limits gate speed.  The simplest estimate for the maximum allowed drive amplitude of the CR drive can be found by equating the control-qubit effective drive amplitude given by $\Omega_{A,0}$ and drive detuning $\Delta_{AB}$, which results in $\lambda \sim 1$, see Eq.~\eqref{lambda}. Then, using Eq.~\eqref{Omega_10_11_nAonly}, we find the corresponding speed limit for the gate duration
\begin{equation}\label{t_min}
    t_{\rm fsl} =  \frac{n^A_{01}}{2 \left|\bra{10}\hat{n}_A \ket{11} - \bra{00}\hat{n}_A \ket{01} \right|} \frac{2\pi}{\Delta_{AB}}\,,
\end{equation}
which we refer to as the fundamental speed limit. A similar criteria based on the off-resonant drive of the control qubit but resulting in $\lambda \sim 1/2$ was used in Refs.~\cite{deGroot2010, deGroot2012} to estimate the maximum possible gate rate. The detuning between qubit frequencies $\Delta_{AB}$ is not restricted for small values for fluxoniums unlike transmons, where the choice of $\Delta_{AB}$ is often affected by the weak anharmonicity~\cite{Tripathi2019}.

We note that Eq.~\eqref{Omega_10_11_nAonly} used to derive Eq.~\eqref{t_min} is valid only in the limit of small $\lambda$.
 When terms of higher order in the drive amplitude are accounted for,  Eq.~\eqref{Omega_10_11_nAonly} is modified and $\Omega_{10-11}$ saturates as a function of $f$ at larger $f$~\cite{Tripathi2019}. Therefore, for $\lambda\sim 1$, the actual value of $\Omega_{10-11}$ is lower than that given by Eq.~\ref{Omega_10_11_nAonly} and the corresponding gate is longer than  $t_{\rm fsl}$. In practice,  experimentally implemented CR gates are typically at least three times the speed limit based on  Eq.~\eqref{t_min} modified for an appropriate system. 
Nevertheless, a quantum speed limit that is close to Eq.~\eqref{t_min} was discovered numerically in the analysis of Ref.~\cite{Kirchhoff2018} for a system of coupled transmons with Eq.~\eqref{t_min} modified for transmons. Close to that speed limit, however, very complex optimal-control pulses were necessary, which were found assuming unconstrained amplitude piecewise-constant controls with a high sampling rate of 0.1 ns. 

In transmons, because of their weak anharmonicity, error due to coherent leakage into the $\ket{1}-\ket{2}$ transition of the control qubit can be significant and give an additional speed-limit restriction. Even though fluxoniums are strongly anharmonic, the corresponding charge matrix element $n^\alpha_{12}$ and thus the resonance Rabi frequency are typically a few times larger than than those for the main $\ket{0}-\ket{1}$ transition, see Table~\ref{Table-params}, which can still contribute to the coherent control error. To elaborate on this issue,
 we first define the dimensionless drive amplitude for the $\ket{1_A}-\ket{2_A}$ transition similarly to Eq.~\eqref{lambda}:
\begin{equation}
    \lambda_{1-2} = \frac{\Omega_{A, 0}^{12}}{\Delta^{12}_{AB}}\,.
\end{equation}
Here $\Omega_{A, 0}^{12} = 2f n^A_{12}$  and $\Delta^{12}_{AB} = \omega^A_{12} - \omega^B_{01}$ are the corresponding single-qubit resonance Rabi frequency and detuning between the transition and drive frequencies.
For parameters of Table~\ref{Table-params}, we find $\lambda_{1-2}/\lambda\approx 0.73 < 1$. Therefore, gate speed is primarily limited by the main $\ket{0_A}-\ket{1_A}$ transition of qubit $A$ as discussed above, although leakage error can be comparable to the error due to rotation of the control qubit, see Sec.~\ref{sec-error-budget} below for the discussion of error budget.

\section{Gate fidelity}\label{sec-fidelity}

\subsection{Unitary dynamics}\label{sec-unitary}

\begin{figure}
 \includegraphics[width=\columnwidth]{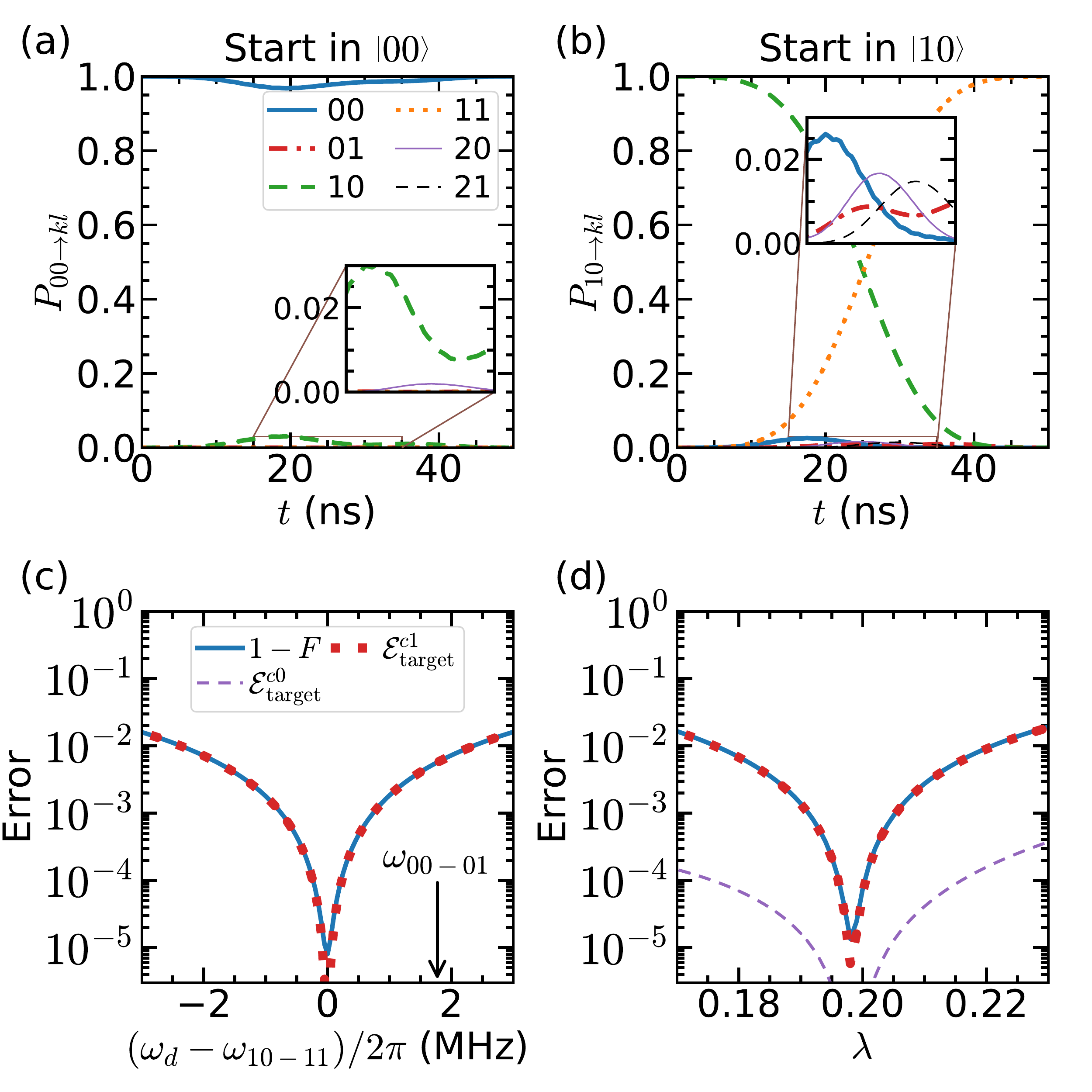}\caption{(a), (b) Unitary time evolution of the populations of various two-qubit states during gate operation for the initial states $\ket{00}$ (a) and $\ket{10}$ (b). The gate is optimized  over drive frequency and two drive amplitudes at fixed $J_C/h = 350$ MHz and $t_{\rm gate}=50$ ns, resulting in $F_{\rm coherent}>99.998\%$. (c), (d) Coherent gate error (blue solid lines) and target-rotation errors for control qubit in $\ket{0}$ (dashed purple lines) and in $\ket{1}$ (dotted red lines) vs drive frequency (c) and dimensionless drive amplitude (d) around their optimal values.}\label{fig_tdomain}
\end{figure}

In numerical simulations of gate dynamics, we use Gaussian envelopes
\begin{equation}\label{pulse_shape}
    f(t) \propto \exp\left[-\frac{(t-t_{\rm gate}/2)^2}{2\sigma^2}\right] - \exp\left[-\frac{(t_{\rm gate}/2)^2}{2\sigma^2}\right]
\end{equation}
for the microwave-drive term~\eqref{drive}.
Here $t_{\rm gate}$ is the gate duration, $0<t<t_{\rm gate}$, and $\sigma = t_{\rm gate}/4$. We have verified that substituting these envelopes with Gaussian flat-top pulses does not significantly affect gate error. For given pulse and other parameters, we first find the unitary evolution operator in the 25-levels Hilbert space composed of five levels coming from each qubit and then project it into the computational subspace to obtain a $4\times 4$ matrix $\hat{U}_{\rm sim}$. To compare it with the ideal \textsc{cnot} operation defined by 
\begin{equation}\label{cnot_def}
 \hat{U}_{\textsc{cnot}} = 
 \begin{pmatrix}
  1 & 0 & 0 & 0\\
  0 & 1 & 0 & 0 \\
  0 & 0 & 0 & 1 \\
  0 & 0 & 1 & 0
 \end{pmatrix}\,,
\end{equation}
we apply additional single-qubit $Z$ rotations both before and after the gate, see Appendix~\ref{sec-single-z}. In an experimental setting, such rotations can be performed instantly in  software and do not contribute to gate error~\cite{McKay2017}. When $\hat{U}_{\rm sim}$ has the same amplitudes of matrix elements as $\hat{U}_{\textsc{cnot}}$, these $Z$ rotations reduce $\hat{U}_{\rm sim}$ to $\hat{U}_{\textsc{cnot}}$ exactly, so any phase error in $\hat{U}_{\rm sim}$ does not contribute to gate infidelity. 

Let $\hat{U}$ be the simulated gate operator $\hat{U}_{\rm sim}$ modified by the additional $Z$ rotations. To calculate the coherent gate fidelity, we use the standard expression~\cite{Pedersen2007}
\begin{equation}\label{fidelity-definition}
 F_{\rm coherent} = \frac{{\rm Tr}\left(\hat{U}^\dagger \hat{U}\right) + \left|{{\rm Tr}\left(\hat{U}_{\textsc{cnot}}^\dagger \hat{U}\right)}\right|^2}{20}\,.
\end{equation}
We optimize it numerically over the drive frequency, the overall drive amplitude, and the parameter $\eta$, see Eq.~\ref{drive}, which is equivalent to optimizing over the two drive amplitudes for drives applied to the control and target qubits. As starting values in the optimization procedure for a given $t_{\rm gate}$, we use $\omega_d = \omega_{10-11}$, take $\eta$ from Eq.~\eqref{eta}, and calculate the overall amplitude prefactor from Eq.~\eqref{Omega_10_11_nAonly} assuming $\Omega_{10-11} = \pi / t_{\rm gate}$ with an extra rescaling to account for the Gaussian pulse shape~\eqref{pulse_shape}. 

In Figs.~\ref{fig_tdomain}(a) and \ref{fig_tdomain}(b), we illustrate such an optimized gate in time domain for $t_{\rm gate}=50$ ns and $J_C/h=350$ MHz by plotting populations of several two-qubit states for two initial states. The coherent gate fidelity~\eqref{fidelity-definition} for these parameters is greater than $99.998\%$. In insets, we show intermediate populations of some of the states that should have zero occupation probabilities in an ideal gate operation. Having $P_{00\to 10}\ne 0$ in Fig.~\ref{fig_tdomain}(a) and $P_{10\to 00}\ne 0$ in Fig.~\ref{fig_tdomain}(b), where $P_{kl \to k'l'}$ is the population of state $\ket{k'l'}$ for the initial state $\ket{kl}$, implies a small chance of spurious control-qubit rotations, and having $P_{10\to 20}\ne 0$ and $P_{10\to 21}\ne 0$ in Fig.~\ref{fig_tdomain}(b) illustrates  leakage to noncomputational states. We emphasize that all such control-qubit rotations and leakage probabilities are very small at the end of the pulse with a more detailed analysis of the error budget given in Sec.~\ref{sec-error-budget}.

To check the stability of this optimized gate operation with respect to variations of pulse parameters, we plot the coherent gate error vs drive frequency detuning $\omega_d-\omega_{10-11}$  and vs dimensionless drive amplitude $\lambda$  around   their optimal values, see blue solid lines in Figs.~\ref{fig_tdomain}(c) and \ref{fig_tdomain}(d).  We  observe that the coherent fidelity stays above $99.99\%$ in the frequency window of about 500 kHz and in the drive-amplitude window of about 2\% of the optimal drive amplitude, which is readily achievable in experiments. 

We emphasize that the optimal value of $\omega_d$ is close to the frequency of the bright (driven) transition $\omega_{10-11}$ rather than to the frequency of the darkened transition $\omega_{00-01}$, which is shown a vertical arrow in Fig.~\ref{fig_tdomain}(c). This way, a controlled rotation of the target qubit is activated by an on-resonance drive, which reduces effects of static $ZZ$ interaction~\eqref{xi_zz}. To illustrate a high-fidelity gate in a system with strong $ZZ$, here we have chosen system parameters with a $ZZ$ rate of $|\xi_{ZZ}|/2\pi \approx 2$ MHz, see the distance between $\omega_{10-11}$ and $\omega_{00-01}$ in Fig.~\ref{fig_tdomain}(c).  In comparison, in simpler schemes to implement a CR gate with a single CR drive followed by additional rotations of the target qubit~\cite{Chow2011}, an ideal operation implies rotations of the target qubit for both states of the control. This condition is hard to achieve when static $ZZ$ coupling is strong, which makes it impossible to drive the target qubit in resonance for both states of the control qubit. In this situation, the $ZZ$ term can be mitigated by an echo sequence~\cite{Corcoles2013, Takita2016, Sheldon2016b, Malekakhlagh2020, Sundaresan2020}, which increases  gate duration. Therefore, the SD scheme has a definite advantage in systems with stronger $ZZ$. Another advantage of our technique is that the error due to extra dynamical phase induced by $ZZ$ interaction does not contribute to gate infidelity because of additional single-qubit $Z$ rotations both before and after the gate operation as described in Appendix~\ref{sec-single-z}. 

We next study coherent gate fidelity as a function of the pulse duration $t_{\rm gate}$ at fixed $J_C/h=350$ MHz and  as a function of $J_C/h$ at fixed $t_{\rm gate}=50$ ns with the gate optimizations performed separately for each value of  $t_{\rm gate}$ or $J_C/h$. The resulting error $1-F_{\rm coherent}$ is shown by the solid blue lines in Figs.~\ref{Fig-fidelity-tgdep} and \ref{Fig-fidelity-jdep}, where parameters of Fig.~\ref{fig_tdomain} are indicated by squares and vertical arrows. We discuss the observed behavior of coherent error in more detail together with  the error budget in Sec.~\ref{sec-error-budget}.

\subsection{Coherent error budget}\label{sec-error-budget}

Here we define several contributions to $1-F_{\rm coherent}$, relating them to those transition probabilities $P_{kl\to k'l'}(t_{\rm gate})$ that should be zero for an ideal gate operation. For brevity, we omit the argument $t_{\rm gate}$ in these probabilities below. To derive the error-budget contributions, we first notice that $|\langle k'l' |\hat{U}|kl\rangle| = \sqrt{P_{kl \to k'l'}}$ when both states $\ket{kl}$ and $\ket{k'l'}$ are in the computational subspace. Then, we use Eq.~(\ref{fidelity-definition}) and  the probability conservation law $\sum_{k'l'}P_{kl \to k'l'} = 1$, where the sum runs over all the two-qubit states, including noncomputational ones, to express $1-F_{\rm coherent}$ in terms of only those $P_{kl\to k'l'}$ that should be zero. Linearizing the resulting expression, we write 
\begin{equation}\label{error_budget}
    1 - F_{\rm coherent} = \mathcal{E}^{c0}_{\rm target} + \mathcal{E}^{c1}_{\rm target} + \mathcal{E}_{\rm control} + \mathcal{E}_{\rm leakage}
\end{equation}
with the individual terms explained below.

The first two terms of Eq.~\eqref{error_budget} represent the errors in the rotation of the target qubit for two states of the control qubit.
When the control qubit is in its ground state, we have
\begin{equation}\label{error_target0}
    \mathcal{E}^{c0}_{\rm target} = \frac 15 \left(P_{00\to 01} + P_{01\to 00}\right)\,,
\end{equation}
implying error due to rotation of the target qubit when it should remain idle.
When the control qubit is in its state $\ket{1}$, we similarly find
\begin{equation}\label{error_target1}
    \mathcal{E}^{c1}_{\rm target} = \frac 15 \left(P_{10\to 10} + P_{11\to 11}\right)\,,
\end{equation}
implying the error due to rotation of the target qubit on the Bloch sphere by an angle different from $\pi$. These two errors are shown by dashed purple and dotted red lines in Figs.~\ref{fig_tdomain}(c), \ref{fig_tdomain}(d), \ref{Fig-fidelity-tgdep}, and \ref{Fig-fidelity-jdep}.  We observe that $\mathcal{E}^{c1}_{\rm target}$ gives the dominant contribution to $1-F_{\rm coherent}$ when the pulse frequency or overall amplitude are slightly detuned from their optimal values, see Figs.~\ref{fig_tdomain}(c) and \ref{fig_tdomain}(d). At the same time, $\mathcal{E}^{c0}_{\rm target}$ remains unimportant in these cases since the ratio of the two drive amplitudes is kept fixed, which enforces the SD condition~\eqref{sd_condition}. When $\eta$ is also detuned from its optimal value (not shown in plots), the $\ket{00}-\ket{01}$ transition is no longer darkened, leading to a larger $\mathcal{E}^{c0}_{\rm target}$. A small increase in $\mathcal{E}^{c0}_{\rm target}$ with deviation of the drive amplitude $\lambda$ in Fig.~\ref{fig_tdomain}(d) is associated with nonlinear corrections, which are absent in the SD condition~\eqref{sd_condition}, but lead to a small change of the optimal $\eta$. For optimized pulses for a given $t_{\rm gate}$, we find that $\mathcal{E}^{c0}_{\rm target}$ and $\mathcal{E}^{c1}_{\rm target}$ have similar values at short $t_{\rm gate}$  with their sum being the dominant contribution in $1-F_{\rm coherent}$, see Fig.~\ref{Fig-fidelity-tgdep}. At long $t_{\rm gate}$ or large $J_C/h$, we find that $\mathcal{E}^{c1}_{\rm target}$ increases and eventually determines $1-F_{\rm coherent}$. 

The third term in Eq.~\eqref{error_budget} is the error due to transitions in the control qubit within the computational subspace:
\begin{equation}\label{error_control}
    \mathcal{E}_{\rm control} = \frac 15 \sum_{k, l, l'=0, 1} P_{kl \to \bar{k}l'}\,,
\end{equation}
where $\bar{0}=1$ and vice versa. This error never dominates $1-F_{\rm coherent}$ for data points presented in this paper and is absent on the plots. 

Finally, the last term in Eq.~\eqref{error_budget} describes the coherent leakage to higher noncomputational levels and is given by
\begin{equation}\label{error_leakage}
    \mathcal{E}_{\rm leakage} = 1 - \frac 14 {\rm Tr}\left(\hat{U}^\dagger \hat{U}\right)\,.
\end{equation}
It is shown by green dash-dot lines in Figs.~\ref{Fig-fidelity-tgdep} and \ref{Fig-fidelity-jdep}. This leakage error is the dominant contribution to $1-F_{\rm coherent}$ at intermediate $t_{\rm gate}$ and is the reason of its local minimum  at $t_{\rm gate}\approx 32$ ns. 
We observe that $\mathcal{E}_{\rm target}^{c0}$, $\mathcal{E}_{\rm target}^{c1}$, and $\mathcal{E}_{\rm leakage}$ have discontinuities as a function of $J_C/h$ at $J_C/h\approx 210$ and $\approx 240$ MHz, which is accompanied by kinks in $1-F_{\rm coherent}$, see Fig.~\ref{Fig-fidelity-jdep}. This behaviour is explained by the crossing of different local minima of $1-F_{\rm coherent}$ at these values of $J_C/h$. A different optimization protocol, e.g., the one that minimizes $\mathcal{E}_{\rm leakage}$ rather than $1-F_{\rm coherent}$, can result in smoother behavior of the total error.

We notice that $1-F_{\rm coherent}$ vs $t_{\rm gate}$ in Fig.~\ref{Fig-fidelity-tgdep} crosses the 0.01 threshold at around 27 ns. At the same time, the fundamental speed limit~\eqref{t_min} is less than 10 ns for these parameters, which can be easily understood by noticing that $\lambda \lesssim 0.2$ at $t_{\rm gate} = 50$ ns, see Fig.~\ref{fig_tdomain}(d), and that Eq.~\eqref{t_min} is based on the $\lambda \sim 1$ criterion. This discrepancy between the shortest possible $t_{\rm gate}$ in a realistic gate and $t_{\rm fsl}$ given by Eq.~\eqref{t_min} is in line with our previous reasoning based on effects that are nonlinear in the drive amplitude, see text below Eq.~\eqref{t_min}. To approach the limit given by Eq.~\eqref{t_min}, optimal-control pulses, which have more complicated shapes than the Gaussian envelope~\eqref{pulse_shape}, are necessary~\cite{Kirchhoff2018}.

\subsection{Dissipation effects}\label{sec-dissipation}

\begin{figure}[t]
\includegraphics[width=\columnwidth]{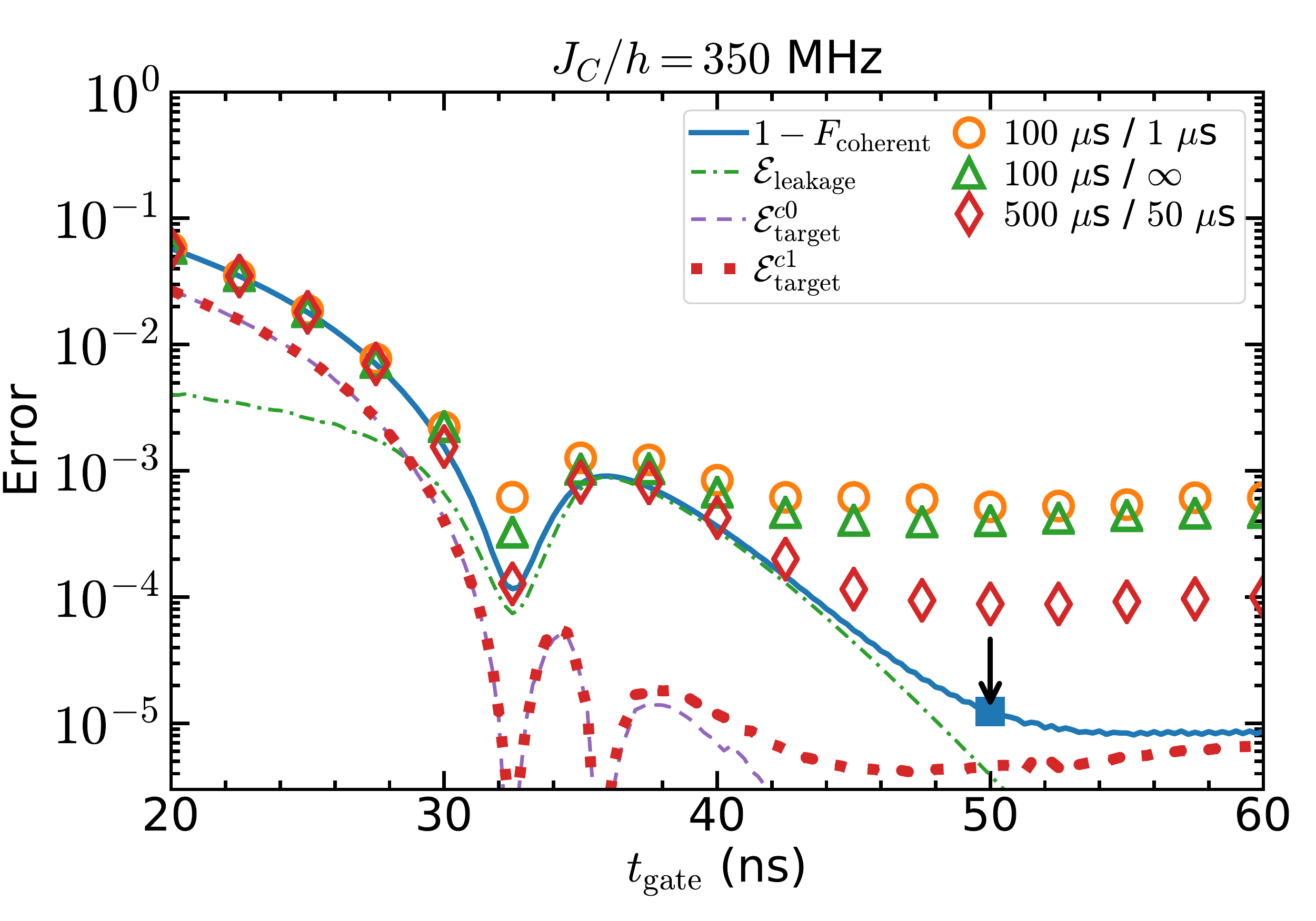}\caption{ Gate error vs $t_{\rm gate}$ at $J_C/h=350$ MHz with pulse parameters optimized for each value of $t_{\rm gate}$. Lines show the total coherent error (blue solid line), leakage error (green dash-dot line), and target-qubit rotation errors for the control qubit in states $\ket{0}$ (purple dashed line) and in $\ket{1}$ (red dotted line). The square marker and vertical arrow point at the parameters of Figs.~\ref{fig_tdomain}(a) and \ref{fig_tdomain}(b). Empty markers show gate error in the presence of relaxation processes and relaxation-limited dephasing for $T_1^{0-1}=100$ $\mu$s and $T_1^{1-2}=1$ $\mu$s (circles), for $T_1^{0-1}=100$ $\mu$s and $T_1^{1-2}=\infty$ (triangles), and for $T_1^{0-1}=500$ $\mu$s and $T_1^{1-2}=50$ $\mu$s (diamonds). }\label{Fig-fidelity-tgdep}
\end{figure}

\begin{figure}[t]
\includegraphics[width=\columnwidth]{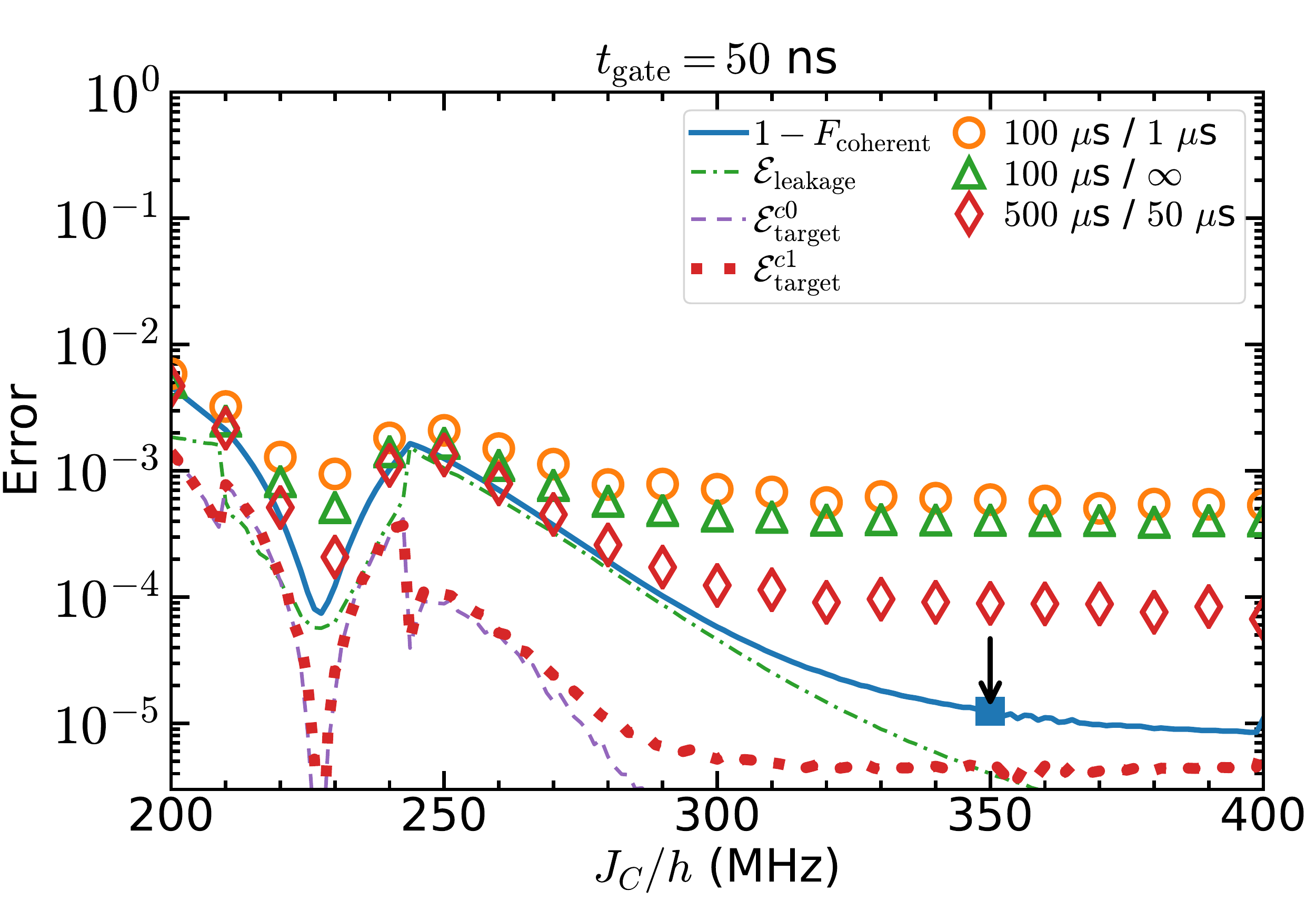}\caption{
Gate error vs $J_C/h$ at $t_{\rm gate}=50$ ns  with pulse parameters optimized for each value of $J_C/h$. Lines show the total coherent error (blue solid line), leakage error (green dash-dot line), and target-qubit rotation errors for the control qubit in states $\ket{0}$ (purple dashed line) and in $\ket{1}$ (red dotted line). The square marker and vertical arrow point at the parameters of Figs.~\ref{fig_tdomain}(a) and \ref{fig_tdomain}(b). Empty markers show gate error in the presence of relaxation processes and relaxation-limited dephasing for $T_1^{0-1}=100$ $\mu$s and $T_1^{1-2}=1$ $\mu$s (circles), for $T_1^{0-1}=100$ $\mu$s and $T_1^{1-2}=\infty$ (triangles), and for $T_1^{0-1}=500$ $\mu$s and $T_1^{1-2}=50$ $\mu$s (diamonds).
}\label{Fig-fidelity-jdep}
\end{figure}

To complete our analysis of gate operation, we discuss the error due to incoherent effects. We focus on qubit relaxation processes and assume that they give the dominant contribution to the loss of coherence, resulting in $T_2 = 2T_1$ relation for the coherence ($T_2$) and relaxation ($T_1$) times of all the relevant transitions.  Since the system is mostly staying in the computational subspace with only small intermediate excitation of qubit second excited states, see Figs.~\ref{fig_tdomain}(a) and \ref{fig_tdomain}(b), the dominant error is expected to come from incoherent effects in $\ket{0_\alpha}-\ket{1_\alpha}$ transitions of both qubits. We denote the corresponding relaxation time by $T_1^{0-1}$ with the assumption that it is the same in both qubits. Because the lifetime of the second excited states can be much shorter than $T_1^{0-1}$, we also account for the relaxation in  $\ket{1_\alpha}-\ket{2_\alpha}$ transitions with the corresponding time $T_1^{1-2}$ for both qubits. 
We follow the procedure outlined in Ref.~\cite{Nesterov2021} to simulate gate dynamics by solving the full master equation, to find the resulting $16 \times 16$ $\chi$ matrix, and then to calculate gate fidelity by comparing this matrix to its ideal value. We implement this procedure for pulse parameters that are optimized for coherent error~\eqref{fidelity-definition}.  

We first consider a suboptimal value $T_1^{0-1}=100$ $\mu$s for the main qubit transitions and a very short relaxation time  $T_1^{1-2}=1$ $\mu$s of the $\ket{1_\alpha}-\ket{2_\alpha}$ transitions. We find that these parameters still result in gate error below $10^{-3}$ for most of the data points, see circles in Figs.~\ref{Fig-fidelity-tgdep} and \ref{Fig-fidelity-jdep}. To estimate the contribution of relaxation specifically in $\ket{0_\alpha}-\ket{1_\alpha}$ transitions, we then remove the  collapse operators corresponding to the $\ket{2_\alpha}-\ket{1_\alpha}$ relaxation from the master equation and compare the two results. We find that gate error for simulations without relaxation in $\ket{1_\alpha}-\ket{2_\alpha}$ transitions is  slightly smaller, see triangles in Figs.~\ref{Fig-fidelity-tgdep} and \ref{Fig-fidelity-jdep} labeled as $T_1^{1-2}=\infty$. Even though $T_1^{1-2}$ is only 1 $\mu$s, the contribution of relaxation in $\ket{1_\alpha}-\ket{2_\alpha}$ transitions to gate error is below $0.1\%$, which agrees with very low excitation probability of noncomputational states during the gate operation, see Figs.~\ref{fig_tdomain}(a) and \ref{fig_tdomain}(b). Finally, we present the results for devices with $T_1^{0-1}=500$ $\mu$s and $T_1^{1-2}=50$ $\mu$s, see diamond-shape markers in Figs.~\ref{Fig-fidelity-tgdep} and \ref{Fig-fidelity-jdep}. These longer relaxation times, which are realistically achievable in  modern fluxonium devices~\cite{Nguyen2019, Somoroff2021}, reduce gate error down to $10^{-4}$ values.

\section{Conclusions}\label{sec-conclusions}

In conclusion, we have investigated a microwave-activated two-qubit gate scheme for fluxonium circuits, which is based on selective darkening of a transition in the computational subspace~\cite{deGroot2010, deGroot2012}. The scheme is facilitated by the cross-resonance effect~\cite{Paraoanu2006, Rigetti2010, Chow2011} and leads to  high-fidelity \textsc{cnot} gates even for strong $ZZ$ coupling and large detuning between qubit frequencies. The gate fidelity in excess of $99.99\%$ evaluated for the unitary dynamics is achievable for a  basic shape of the microwave radiation and does not require  complicated pulse sequences or special arrangement of qubit energy levels. The population of higher excited states remains low during gate pulses, so the gate performance is nearly unaffected by the relaxation processes from energy levels outside of the computational subspace. Even for a short lifetime of 1 $\mu$s of the second excited state, the contribution to the gate error  remains below 0.1\%. 

The optimized gate operation was analyzed in this paper for a device with a relatively strong $ZZ$ coupling, which was   $\xi_{ZZ}/2\pi \approx 2$ MHz in the case discussed in Fig.~\ref{fig_tdomain}. This example illustrates an excellent resilience of the SD gates against spurious $ZZ$ coupling. While this specific \textsc{cnot} gate works well for this value of $\xi_{ZZ}$, it is generally preferred to have processors with much smaller values of the $ZZ$  crosstalk during   single-qubit gates and while idling.
The magnitude of $ZZ$ coupling can be greatly reduced by applying an always-on off-resonance microwave drive, which was demonstrated for fluxoniums~\cite{Xiong2022} and transmons~\cite{Mitchell2021, Wei2022}. Additional techniques to mitigate this crosstalk are based on using a multipath coupling scheme~\cite{Kandala2021, Mundada2019}  and/or a tunable coupler~\cite{Mundada2019, Yan2018}, which will likely become an ultimate scalable solution for a multiqubit processor.
Without such techniques, for fluxonium parameters used in our analysis, a 50 ns-long high-fidelity gate is possible for a smaller value of $J_C/h$ with $\xi_{ZZ}/2\pi < 1$ MHz, see Fig.~\ref{Fig-fidelity-jdep}. Because gate rate~\eqref{Omega_10_11_nAonly} scales with $J_C$ linearly, while $\xi_{ZZ}$ is quadratic in $J_C$, an additional reduction of $J_C/h$ by a factor of 2 increases $t_{\rm gate}$ to 100 ns, but reduces $\xi_{ZZ}$ by an extra factor of 4. We also note that here we have performed a very basic optimization procedure of microwave pulses, which can be greatly improved with optimal control to allow a fast gate with even smaller $J_C/h$ or $\xi_{ZZ}$. 

In transmons, the rate of the cross-resonance gate is maximized when two qubits are in the straddling regime, so the detuning between qubit frequencies is smaller than their anharmonicity~\cite{Tripathi2019}. In addition, in a multiqubit transmon processor, an extra care is required to tune frequencies to reduce spectator errors~\cite{Sundaresan2020}. These requirements of simultaneously having small detunings  and avoiding frequency collisions, including transitions to the second excited states, lead to a complicated frequency-allocation procedure for transmon-based processors~\cite{Morvan2022}. Modern fabrication techniques result in 1\% imprecision in transmon frequencies~\cite{Kreikebaum2020}, which is insufficient to obtain a high yield of devices that satisfy frequency conditions unless an additional postfabrication tuning is performed~\cite{Zhang2022, Hertzberg2021}. 

In comparison, frequency requirements for fluxoniums are much less stringent, which guarantees a better fabrication yield of fluxonium-based processors. The road map for a scalable fluxonium-based processor, including frequency allocation and estimates for the fabrication yield, has been recently presented in Ref.~\cite{Nguyen2022}. There, it was argued that the two-qubit gate realized via the CR effect is a viable solution for a scalable design for such a processor. In addition, it was shown that a combination of capacitive and inductive couplings can effectively suppress  the static $ZZ$ rate, while maintaining high-fidelity CR gates. 

Finally, another attractive feature of the fluxonium is that its low frequency implies a long coherence time, which currently exceeds 1 ms in best devices~\cite{Somoroff2021}. We have shown that comparable qubit lifetimes together with realistic relaxation time of 50 $\mu$s of the second excited states result in gate error within the $10^{-4}$ range. We note that the fluxonium frequency in Ref.~\cite{Somoroff2021} was 163 MHz, while devices with higher qubit frequencies are preferred for the realization of a high-fidelity SD scheme to increase hybridization of qubit states in a system with capacitive coupling. Thus, simulations in this paper were performed for two qubits with frequencies in the 500 MHz -- 1 GHz range. Generally, higher-frequency fluxoniums have proportionally lower coherence times. 
However, much less community effort has been devoted to improving fluxonium devices in comparison to transmons. In particular,  suboptimal fabrication procedure and antenna design were used in high-coherence fluxonium devices of Refs.~\cite{Nguyen2019, Somoroff2021}, resulting in effective dielectric loss tangents that are an order of magnitude larger than in the best 3D transmons~\cite{Wang2015}. Given recent advances in materials and fabrications of 2D transmons~\cite{Place2021, Wang2022a}, the fluxonium coherence time can be pushed up significantly both in planar and 3D geometries.
These arguments indicate that the fluxonium is an excellent candidate for a scalable processor, and the SD gate scheme is suitable for its realization.
 
\begin{acknowledgements}

We would like to thank Long Nguyen, Quentin Ficheux, Haonan Xiong, Lo\"ick Le Guevel, Ebru Dogan, and Dario Rosenstock for stimulating discussions. We acknowledge the support from  ARO-LPS HiPS program (grant No. W911NF-18-1-0146). V.E.M. and M.G.V acknowledge the Faculty Research Award from Google and fruitful conversations with the members of the Google Quantum AI team.  We used the QuTiP software package~\cite{Johansson2012, Johansson2013}  and performed computations using resources and assistance of the UW-Madison Center For High Throughput Computing (CHTC) in the Department of Computer Sciences. The CHTC is supported by UW-Madison, the Advanced Computing Initiative, the Wisconsin Alumni Research Foundation, the Wisconsin Institutes for Discovery, and the National Science Foundation.

\end{acknowledgements}

\begin{widetext}
\appendix
\section{Perturbation theory}\label{sec-perturbation}

Here we develop approximate expressions for the cross matrix elements of the charge operators based on first-order perturbation theory. Among an infinite number of all the contributions that are formally linear in $J_C$, we aim at keeping only those whose magnitude is not very small. To this end, we truncate the Hilbert space and account only for contributions coming from hybridization within the computational subspace and only with those noncomputational levels where one of the qubits is in its second or third excited states.
We first define
\begin{equation}
   \hbar V_{kl, k'l'} = \bra{kk'} \hat{V} \ket{ll'} = J_C n^A_{kl}n^B_{k'l'}\,,
\end{equation}
where
\begin{equation}
    n^\alpha_{kl} = -i\bra{k_\alpha}\hat{n}_\alpha \ket{l_\alpha}\,.
\end{equation}
Keeping only relevant terms, we find
\begin{subequations}
\begin{equation}
\begin{aligned}
\ket{00} &\approx \ket{0_A}\ket{0_B} - \frac{V_{11, 00}}{\omega^A_{01} + \omega^B_{01}}\ket{1_A}\ket{1_B} - \frac{V_{31, 00}}{\omega^A_{03} + \omega^B_{01}}\ket{3_A}\ket{1_B} - \frac{V_{13, 00}}{\omega^A_{01} + \omega^B_{03}}\ket{1_A}\ket{3_B} + \ldots\,,
\end{aligned}
\end{equation}
\begin{equation}
\begin{aligned}
\ket{11} &\approx \ket{1_A}\ket{1_B} + \frac{V_{00, 11}}{\omega^A_{01} + \omega^B_{01}}\ket{0_A}\ket{0_B}  
- \frac{V_{20, 11}}{\omega^A_{12} - \omega^B_{01}}\ket{2_A}\ket{0_B} + \frac{V_{02, 11}}{\omega^A_{01} - \omega^B_{12}}\ket{0_A}\ket{2_B} + \ldots\,,
\end{aligned}
\end{equation}
\begin{equation}
\begin{aligned}
\ket{01} &\approx \ket{0_A}\ket{1_B} - \frac{V_{10, 01}}{\omega^A_{01} - \omega^B_{01}}\ket{1_A}\ket{0_B}  - \frac{V_{12, 01}}{\omega^A_{01} + \omega^B_{12}}\ket{1_A}\ket{2_B} - \frac{V_{30, 01}}{\omega^A_{03} - \omega^B_{01}}\ket{3_A}\ket{0_B} + \ldots\,,
\end{aligned}
\end{equation}
and
\begin{equation}
\begin{aligned}
\ket{10} &\approx \ket{1_A}\ket{0_B} + \frac{V_{01, 10}}{\omega^A_{01} - \omega^B_{01}}\ket{0_A}\ket{1_B}  - \frac{V_{21, 10}}{\omega^A_{12} + \omega^B_{01}}\ket{2_A}\ket{1_B} + \frac{V_{03, 10}}{\omega^A_{01} - \omega^B_{03}}\ket{0_A}\ket{3_B} + \ldots\,.
\end{aligned}
\end{equation}
\end{subequations}
Therefore, we find the following expressions for the cross matrix elements of $\hat{n}_A$ (e.g., for transitions of qubit $B$):
\begin{subequations}
\begin{equation}
    \begin{aligned}
   \bra{00} &\hat{n}_A \ket{01} \approx  -2i\frac{J_C}{\hbar}
   n^B_{01}\left[ \frac{\left(n^A_{01}\right)^2\omega^A_{01}}{ \left(\omega^A_{01}\right)^2 - \left(\omega^B_{01}\right)^2} +  \frac{\left(n^A_{03}\right)^2\omega^A_{03}}{ \left(\omega^A_{03}\right)^2 - \left(\omega^B_{01}\right)^2}\right]
    \end{aligned}
\end{equation}
and
\begin{equation}
    \begin{aligned}
   \bra{10} &\hat{n}_A \ket{11} \approx  2i\frac{J_C}{\hbar}
   n^B_{01}\left[ \frac{\left(n^A_{01}\right)^2\omega^A_{01}}{ \left(\omega^A_{01}\right)^2 - \left(\omega^B_{01}\right)^2} -  \frac{\left(n^A_{12}\right)^2\omega^A_{12}}{ \left(\omega^A_{12}\right)^2 - \left(\omega^B_{01}\right)^2}\right]\,.
    \end{aligned}
\end{equation}
\end{subequations}
Similarly, for $\hat{n}_B$ we find
\begin{subequations}
\begin{equation}
    \begin{aligned}
   \bra{00} &\hat{n}_B \ket{10} \approx  2i\frac{J_C}{\hbar}
   n^A_{01}\left[ \frac{\left(n^B_{01}\right)^2\omega^B_{01}}{ \left(\omega^A_{01}\right)^2 - \left(\omega^B_{01}\right)^2} + \frac{\left(n^B_{03}\right)^2\omega^B_{03}}{\left(\omega^A_{01}\right)^2 -  \left(\omega^B_{03}\right)^2  }\right]
    \end{aligned}
\end{equation}
and
\begin{equation}
    \begin{aligned}
   \bra{01} &\hat{n}_B \ket{11} \approx - 2i\frac{J_C}{\hbar}
n^A_{01}\left[ \frac{\left(n^B_{01}\right)^2\omega^B_{01}}{ \left(\omega^A_{01}\right)^2 - \left(\omega^B_{01}\right)^2} -  \frac{\left(n^B_{12}\right)^2\omega^B_{12}}{ \left(\omega^A_{01}\right)^2 - \left(\omega^B_{12}\right)^2}\right]\,.
    \end{aligned}
\end{equation}
\end{subequations}

\section{Single-qubit $Z$ rotations}\label{sec-single-z}

An evolution operator projected into the computational subspace has the form
\begin{equation}
 \hat{U}_{\rm sim} = \begin{pmatrix}
  Xe^{i\phi_{00}} & x & x & x \\
  x &  Xe^{i\phi_{01}} & x & x \\
  x & x & x & Xe^{i\phi_a} \\
  x & x & Xe^{i\phi_b} & x 
 \end{pmatrix}\,,
\end{equation}
where $X$ stands for absolute values of matrix elements that are equal to 1 in the ideal operator and $x$ is used to describe the remaining elements. 
To compare this operator with the ideal one, we use additional single-qubit $Z$ rotations both \emph{before and after} the gate operation. Namely,
\begin{equation}
 \hat{U}_{\rm sim} \rightarrow \hat{U}= e^{-i\phi_{00}}
 \begin{pmatrix}
 1 & 0 & 0 & 0 \\
 0 & e^{i\phi_1} & 0 & 0 \\
 0 & 0 & e^{i\phi_2} & 0 \\
 0 & 0 & 0 & e^{i(\phi_1 + \phi_2)}
 \end{pmatrix}
 \begin{pmatrix}
  Xe^{i\phi_{00}} & x & x & x \\
  x &  Xe^{i\phi_{01}} & x & x \\
  x & x & x & Xe^{i\phi_a} \\
  x & x & Xe^{i\phi_b} & x 
 \end{pmatrix}
 \begin{pmatrix}
 1 & 0 & 0 & 0 \\
 0 & e^{i\phi_3} & 0 & 0 \\
 0 & 0 & 1 & 0 \\
 0 & 0 & 0 & e^{i\phi_3 }
 \end{pmatrix}\,.
\end{equation}
Thus, before the gate operation, we perform additional $Z$ rotation on the target qubit only (phase $\phi_3$), while rotations on both target and control are applied after the operation.
The phases determining these three rotations are given by
\begin{eqnarray}
 \left\{
 \begin{array}{ll}
  \phi_1 &= \frac 12 \left(-\phi_{01} + \phi_a - \phi_b + \phi_{00}\right)\,, \\
  \phi_2 &= \frac 12 \left(\phi_{01} - \phi_a - \phi_b + \phi_{00}\right)\,, \\
  \phi_3 &= \frac 12 \left(-\phi_{01} - \phi_a + \phi_b + \phi_{00}\right)\,\,,
 \end{array}
 \right.
\end{eqnarray}
which nullifies all the phases of important matrix elements (labeled by $X$).

\end{widetext}

\bibliography{literature}

\end{document}